\documentclass[aps,showpacs,showkeys,twocolumn,prl]{revtex4}
\usepackage{amssymb}
\usepackage{natbib}
\usepackage{amsmath}
\usepackage{amsfonts}
\usepackage{graphicx}
\usepackage{mathrsfs}
\usepackage{dcolumn}
\usepackage{bm}
\usepackage{color}
\definecolor{rot}{rgb}{0.75,0.05,0.25}
\definecolor{hellgrau}{gray}{0.5}
\definecolor{blau}{rgb}{0,0,0.7}

\def\Tr{\mbox{Tr}}

\begin{document}

\title{Fluctuation Theorem for Arbitrary Open Quantum Systems}
\author{Michele Campisi, Peter Talkner, Peter H\"anggi}
\affiliation{Institute of Physics, University of Augsburg,
  Universit\"atstr. 1, D-86135 Augsburg, Germany}
\date{\today }

\begin{abstract}
Based on the observation that the thermodynamic equilibrium free energy
of an open quantum system in contact with a thermal environment can
be understood as the difference between  the free energy of the
total system and that of the bare environment, the validity of the
Crooks theorem and of the Jarzynski equality is extended  to open
quantum systems. No restrictions on the nature of the environment or
on the strength of the coupling between system and environment need
to be imposed. This free energy entering the Crooks theorem and the
Jarzynski equality is closely related to the Hamiltonian of mean
force that generalizes the classical statistical mechanical concept
of the potential of mean force.

\end{abstract}
 \maketitle

Since its formulation in 1997 the classical nonequilibrium work
relation by Jarzynski \cite{Jarz97} (now commonly referred to as
Jarzynski Equality)
\begin{equation}
 \langle e^{-\beta w} \rangle = e^{-\beta \Delta F}
 \label{eq:JE}
\end{equation}
kept raising questions and concerns on its range of validity and
applicability. Here $w$ denotes the work performed on a system when
some parameters of this system are changed according to a prescribed
protocol. Initially, the system is supposed to be prepared in a
thermal equilibrium state at the inverse temperature $\beta$. The
brackets $\langle \cdot \rangle$ denote a non-equilibrium average
over many repetitions of this process, running under the same
protocol. According to the Jarzynski equality the average of the
exponentiated negative work is independent of the details of the
protocol and solely determined by the thermal equilibrium free
energy difference $\Delta F$ between the initial equilibrium state and a
hypothetical equilibrium state at the initial temperature and those
parameter values that are reached at the end of the protocol. In the
mentioned paper \cite{Jarz97} the validity of this equality was
demonstrated within a classical statistical approach for isolated
systems which initially are in the required equilibrium state at
inverse temperature $\beta$ and also for classical systems that stay in
{\it weak} contact with a thermal bath during the protocol.

Considerable efforts have been devoted to the development of the
quantum version of Eq. (\ref{eq:JE}), and more generally of the
Crooks fluctuation theorem \cite{Crooks:1999fq} that underlies it,
i.e.,
\begin{equation}
\frac{p_{t_f,t_0}(+w)}{p_{t_0,t_f}(-w)} = e^{\beta (w-\Delta F)} \;,
\label{eq:FT}
\end{equation}
where $p_{t_{f},t_{0}}(w)$ denotes the probability density function
(pdf) of work performed by the parameter changes according to a
protocol running between the initial time $t_{0}$ and final time
$t_{f}$. The pdf of work for the reversed protocol is denoted by
$p_{t_0,t_f}(w)$. All these attempts refer to quantum isolated or
weakly coupled system with Hamiltonian or Markovian dynamics,
respectively
\cite{Tasaki:2000pi,Talkner:2007sf,Mukamel:2003ip,De-Roeck:2004cq,Esposito:2006by,CrooksJSM08}.
The proof for  the validity of Eqs. (\ref{eq:JE}), (\ref{eq:FT}) in
the quantum case with {\it weak coupling} allowing for  an otherwise
general non Markovian dynamics of the open quantum dynamics and
arbitrary force protocols was provided only recently in Ref.
\cite{Talkner:2008mr}.

The applicability of Eqs. (\ref{eq:JE}), (\ref{eq:FT}) to the case
of weak coupling is consistent with the
construction of quantum and classical statistical mechanics which
relies on that assumption. In striking contrast, extending the
methods of statistical mechanics to cases that involve a
non-negligible system-environment interaction presents a major
challenge \cite{HIT_NJP08,Ingold:2008mz}. Addressing this question
is by now becoming more and more pressing, as the advancement of
technology poses us in the position
to investigate experimentally the thermodynamic behavior of
nanosystems operating in the quantum regime, whose reduced sizes
make the system-environment coupling an important issue.

A satisfactory generalization of the applicability of Eq.
(\ref{eq:JE}) for the classical strong coupling regime was put
forward by Chris Jarzynski  himself \cite{Jarzynski:2004pv}; it
should be stressed, however, that the objective of the corresponding
quantum treatment has not been achieved yet. The key tool used in
Ref. \cite{Jarzynski:2004pv} to overcome the difficulties posed by
the presence of strong coupling is the \emph{Hamiltonian of mean
force} $H^*(\Gamma_S,t)$ where $\Gamma_S$ denotes a point in the phase space
of the subsystem of interest. This Hamiltonian of mean force is
defined as the effective Hamiltonian that describes the
Boltzmann-Gibbs equilibrium of the marginal probability density of
the subsystem of interest; it reads
\begin{equation}
\begin{split}
H^*(\Gamma_S;t) & = H_S(\Gamma_S,t) \\
 - \frac{1}{\beta} \ln   &\frac{\int d\Gamma_B
  \exp(-\beta(H_B(\Gamma_B)+H_{SB}(\Gamma_S,\Gamma_B)))}{\int d\Gamma_B \exp(-\beta H_B(\Gamma_B))} \; ,
\end{split}
\label{eq:H*cl}
\end{equation}
where $\Gamma_B$
denotes a point in the phase space of the bath.
The total Hamiltonian of system plus environment is given by
 \begin{equation}
 H(\Gamma_S,\Gamma_B,t)= H_S(\Gamma_S,t)
+H_B(\Gamma_B)+H_{SB}(\Gamma_S,\Gamma_B)
\end{equation}
which is  composed of the Hamiltonian of the isolated system of interest, $H_S(\Gamma_S,t)$
(time-dependent), the bath Hamiltonian $H_B(\Gamma_B)$ and the
interaction Hamiltonian $H_{SB}(\Gamma_S,\Gamma_B)$. This Hamiltonian of
mean force generalizes the concept of the potential of mean force,
see Eq. (\ref{eq:V*cl}) below, that is commonly employed, for
example, in reaction rate theory \cite{HTB1990} and in the study of
implicit solvent models in terms of biomolecular simulations
\cite{Roux19991}.

In the context of quantum rate theory, potentials of mean force have
been determined for the reaction coordinate from path integral
expressions of the partition function of a composed system with the
reaction coordinate confined to a ``centroid''
\cite{HTB1990,PollakTalkner2005}. A direct application of this very
approach to obtain quantum fluctuation theorems for open systems
though is not obvious. Jarzynski, in fact, did emphasize that his
treatment is restricted to the classical case
\cite{Jarzynski:2004pv}. Addressing this problem thus requires a
careful analysis of what one should consider as the system partition
function from which an equilibrium free energy of the system can be
inferred. One could naively take for this partition function   the
bare partition function of the \emph{isolated} system of interest:
This procedure would, however, neglect the prominent  fact that the
interaction with the bath {\it alters} the system properties.
Such a choice, therefore, is generally physically not correct.
Instead of the partition function of the \emph{isolated}
system one has to choose a properly defined partition function,
which embraces the influence of bath on the {\it open} quantum
system.

As we will show in this Letter, the introduction of a proper
partition function for an open system  allows one to prove that both
the Tasaki-Crooks fluctuation theorem and the Jarzynski equality in
fact hold true for general
{\it open} quantum systems, independent of coupling strength. Moreover, from this partition function a
{\it quantum Hamiltonian of mean force} can be inferred which takes
over the role of the Hamiltonian of mean force in classical
statistical mechanics.

\emph{The Argument.}$-$ Consider the Hamiltonian operator
$\hat{H}(t)$ of a quantum system composed of the interacting system
and the bath which we write as
\begin{equation}
\hat{H}(t)=\hat{H}_S(t)+\hat{H}_{SB}+\hat{H}_{B} \; ,
\label{eq:Ham}
\end{equation}
where the system Hamiltonian $\hat{H}_S(t)$ is time-dependent in a
way that results from a pre-specified protocol of system parameter
changes. The interaction Hamiltonian $\hat{H}_{SB}$ and the bath
Hamiltonian $\hat{H}_{B}$ are supposed to be independent of time.
The change of system's parameters can be interpreted as a time
dependent external forcing that is able to perform work on the
system.

The total system is isolated; it therefore obeys the quantum
Tasaki-Crooks Fluctuation Theorem \cite{Tasaki:2000pi,TH2007}:
\begin{equation}
\frac{p_{t_f,t_0}(+w)}{p_{t_0,t_f}(-w)}=\frac{Y(t_f)}{Y(t_0)}e^{\beta w} \; ,
\label{eq:TC}
\end{equation}
with $Y(t)$ being the total partition function,
i.e.,
\begin{equation}
Y(t)=\Tr e^{-\beta (\hat{H}_S(t)+\hat{H}_{SB}+\hat{H}_{B})} \; ,
\label{eq:Y(t)}
\end{equation}
where $\Tr$ denotes the trace over the total system
Hilbert space, and the symbols ${p_{t_f,t_0}(w)}$ and
$p_{t_0,t_f}(w)$ denote the probability densities of doing the work
$w$ when the protocol is run in the forward and backward directions,
respectively. It is important to note that, due to the fact that the
forces solely act on the system, the work performed on the open
system coincides with that done on the total system.

In order to properly define the partition function of the
open quantum system $S$ staying in thermal equilibrium with a bath, we appeal to thermodynamic reasoning. As pointed out in Ref. \cite{Ford:1985it}, the
free energy of the open system of interest is the difference between
the total system free energy and the bare bath free energy:
\begin{equation}
 F_S(t) = F(t) - F_B  \; .
 \label{FS}
\end{equation}
Here $t$ merely specifies the values of the external parameters as
they occur in the course of the protocol at the time $t$. The
function $F_{S}(t)$ satisfies all required properties of a free
energy \cite{HIT_NJP08,Haenggi:2006br,Ingold:2008mz}. Using the
statistical mechanical relation $\beta F = -\ln Z$ between
equilibrium free energy and the partition function, one finds from
Eq. (\ref{FS}) for the partition function of an open quantum system
the result that is well known to those working on strong quantum dissipation 
\cite{Ingold:2008mz,Horhammer,Fundamentals,Haenggi:2006br,Ingold:2002pc,Theo2002,
HaenggiQTransp98,Grabert:1988et,Ford:1985it,Grabert:1984ux,Caldeira:1983aj,FeynStatMech,
Feynman:1963qm}, namely:
\begin{equation}
Z_S(t) = \frac{\Tr
  e^{-\beta(\hat{H}_S(t)+\hat{H}_{SB}+\hat{H}_{B})}}{\Tr_{B} e^{-\beta
    \hat{H}_{B}}}
 = \frac{Y(t)}{Z_B} \;,
\label{eq:Zs=Z-Zb}
\end{equation}
where $\Tr_{B}$ is the trace over the bath Hilbert space, and $Z_B$
is the bare bath partition function, i.e., $Z_B=\Tr_{B} e^{-\beta
\hat{H}_{B}}$, which  is independent of time.

From  Eqs. (\ref{eq:Y(t)}) and (\ref{eq:Zs=Z-Zb}), and the fact that
$Z_B$ does not depend on time $t$, the salient  relation
${Y(t_f)}/Y(t_0)=Z_S(t_f)/{Z_S(t_0)}$ follows. This  quantum result
assumes a form reading just like in in the classical case
\cite{Jarzynski:2004pv}. Therefore, Eq. (\ref{eq:TC}) becomes
\begin{equation}
\frac{p_{t_f,t_0}(+w)}{p_{t_0,t_f}(-w)}=
\frac{Z_S(t_f)}{Z_S(t_0)}e^{\beta w} = e^{\beta(w-\Delta F_S)} \; ,
\label{eq:TCs}
\end{equation}
which states that the ratio of probabilities of work in the backward
and forward protocols is dictated by the equilibrium free energy
difference  of the open quantum system, i.e., $\Delta F_{S} =
F_{S}(t_{f}) -F_{S}(t_{0})$, with $F_{S}(t)$ given by Eq.
(\ref{FS}).

By multiplying Eq. (\ref{eq:TCs}) by $p_{t_0,t_f}(-w) e^{-\beta w}$
and integrating over $w$ in the usual way one obtains  with $\Delta
F_{s}$ the very form (\ref{eq:JE}) of the Jarzynski equality for
open quantum systems, being valid independently of the coupling
strength and the details of the bath. This is in complete analogy
with Jarzynski's classical result, which, therefore, carries over to
the quantum case.
Hence, if a classical
force acts on an open system, the average exponentiated
work $ e^{-\beta w}$ equals the exponentiated
system equilibrium free energy difference, both in  classical and
quantum regimes.

\emph{Remarks.$-$} The system partition function $Z_S(t)$, defined
in Eq. (\ref{eq:Zs=Z-Zb}), is actually the partition function
associated to the quantum Hamiltonian of mean force $\hat{H}^*(t)$,
defined in analogy to the classical Hamiltonian of mean force as:
\begin{equation}
\hat{H}^*(t):= -\frac{1}{\beta}\ln \frac{\Tr_B
  e^{-\beta(\hat{H}_S(t)+\hat{H}_{SB}+\hat{H}_{B})}}{\Tr_{B} e^{-\beta
    \hat{H}_{B}}} \;.
    \label{eq:H*}
\end{equation}
In fact, the partition function $Z_{S}(t)$ can be recast as:
\begin{equation}
Z_{S}(t)= \Tr_S e^{-\beta \hat{H}^*(t)} \;,
\label{eq:Zs(t)}
\end{equation}
 where $\Tr_S$
denotes the trace over the system Hilbert space. Using Eqs.
(\ref{eq:Zs=Z-Zb}) and (\ref{eq:H*}), it follows that
\begin{equation}
Z^{-1}_{S}(t) e^{-\beta \hat{H}^*(t)}= Y^{-1}(t) \Tr_B e^{-\beta
\hat{H}(t)}\;,
\label{rhoS}
\end{equation}
where the right hand side coincides with the reduced density matrix
of the open system in thermal equilibrium with the heat bath. Again,
$t$ merely characterizes those parameter values that occur according
to the protocol at time $t$. It does \emph{not} indicate any dynamics. The
actual time dependent density matrix at time $t$ does {\it not}, in
general, coincide with ${e^{-\beta \hat{H}^*(t)}}/{Z_{S}(t)}$.

We note that  the Hamiltonian of mean force $\hat{H}^{*}(t)$
typically is a complicated operator-valued function not only of the
system's parameters but also of the system-bath coupling strength,
the bath temperature and possibly of other bath parameters. In the
case of weak coupling the contributions from the bath and the
interaction(s) are negligible and the Hamiltonian of mean force
reduces to the bare system Hamiltonian \cite{Talkner:2008mr}.

In the classical limit, the quantum Hamiltonian of mean force
becomes the classical Hamiltonian of mean force in  Eq.
(\ref{eq:H*cl}). This classical expression can  further be simplified for a bath Hamiltonian
consisting of a sum of potential and kinetic energies where the
latter do not depend on positions and for an interaction that is
independent of the bath momenta. Then the mere integration over the
momenta yields identical factors in the numerator and denominator of
Eq. (\ref{eq:H*}), which cancel each other. The remaining term under
the logarithm is then only  a function of the system positions. This
leads to the renormalization of the potential, -- i.e. to the
potential of mean force --, mentioned before:
\begin{equation}
\begin{split}
V^*(x;t) &= V_S(x,t) \\
&\qquad - \frac{1}{\beta} \ln \frac{\int dy
  \exp(-\beta(V_B(y)+V_{SB}(x,y)))}{\int dy \exp(-\beta V_B(y))} \; ,
\end{split}
\label{eq:V*cl}
\end{equation}
whereas the kinetic energy of the system remains unchanged by this
procedure. We note that the given conditions are sufficient but not
necessary in order that the potential of mean force captures the
complete effect of a bath on the equilibrium properties of the
system. Further simplifications result for example for a bath
consisting of a set of harmonic bath oscillators which linearly
couple to phase space functions of the system. Then, classically,
the Hamiltonian of mean force coincides with the bare Hamiltonian of
the system. This is in strong contrast to the behavior of quantum
systems which couple as above, i.e., linearly, to a harmonic heat bath.
In this case, the Hamiltonian operator of mean force  deviates from the bare
system Hamiltonian with respect to the kinetic and the potential
energy \cite{Fundamentals,Grabert:1984ux}.

\emph{Conclusions.$-$} Surprisingly enough, the Crooks theorem and
the Jarzynski equality are valid for open quantum systems
irrespectively of the coupling strength to their thermal environment
and of the particular nature of their environment. These theorems,
hence, are valid for all types of processes in which a classical or
quantum system in contact with a thermal heat bath is driven out of
equilibrium by classical, generally time-dependent forces.

\emph{Acknowledgements. $-$} Financial support by the German
Excellence Initiative via the {\it Nanosystems Initiative Munich}
(NIM) and the Volkswagen Foundation (project I/80424) is gratefully
acknowledged.


\begin{thebibliography}{10}

\bibitem{Jarz97}
C.~Jarzynski,
  Phys. Rev. Lett. {\bf 78}, 2690 (1997).

\bibitem{Crooks:1999fq}
G.~E. Crooks, Phys. Rev. E
  {\bf 60}, 2721 (1999).

\bibitem{Tasaki:2000pi}
H.~Tasaki, preprint arXiv:cond-mat/0009244 (2000).

\bibitem{Talkner:2007sf}
P.~ Talkner, E.~ Lutz, and P.~H\"{a}nggi,
Phys. Rev. E {\bf 75}, 050102(R) (2007). 

\bibitem{Mukamel:2003ip}
S.~Mukamel, Phys. Rev. Lett. {\bf 90}, 170604 (2003).

\bibitem{De-Roeck:2004cq}
W.~De~Roeck and C.~Maes, Phys. Rev. E {\bf 69},
  026115 (2004).

\bibitem{Esposito:2006by}
M.~Esposito and S.~Mukamel, Phys. Rev. E {\bf 73}, 046129 (2006).

\bibitem{CrooksJSM08}
G.~E. Crooks, J. Stat. Mech. (2008) P10023.

\bibitem{Talkner:2008mr}
P.~Talkner, M.~Campisi, and P.~H\"anggi, J. Stat. Mech. (2009) P02025.

\bibitem{HIT_NJP08}
P.~H\"{a}nggi, G.-L. Ingold, and P.~Talkner, New J. of Phys. {\bf 10},
  115008 (2008).

\bibitem{Ingold:2008mz}
G.-L. Ingold, P.~H\"anggi, and P.~Talkner, arXiv:0811.3509 (2008).

\bibitem{Jarzynski:2004pv}
C.~Jarzynski, J. Stat. Mech. (2004) P09005.

\bibitem{HTB1990}
P.~H\"anggi, P.~Talkner, and M.~Borkovec, Rev. Mod. Phys. {\bf 62}, 251 (1990).

\bibitem{Roux19991}
B.~Roux and T.~Simonson, Biophys. Chem. {\bf
  78}, 1 (1999).

\bibitem{PollakTalkner2005}
E.~Pollak and P.~Talkner, Chaos {\bf 15}, 026116 (2005).

\bibitem{TH2007}
P.~Talkner and P.~H\"anggi, J. Phys. A {\bf 40}, F569 (2007).

\bibitem{Ford:1985it}
G. W.~Ford, J. T.~Lewis, and R. F.~O'Connell, Phys. Rev. Lett. {\bf 55}, 2273 (1985).

\bibitem{Horhammer}
C.~H\"orhammer and H.~ B\"uttner, J. ~Stat.~ Phys.~ {\bf 133}, 1161
(2008).

\bibitem{Fundamentals}
P.~H\"anggi and G.~L.~Ingold, Chaos {\bf 15}, 026105 (2005).

\bibitem{Haenggi:2006br}
P.~{H{\"a}nggi} and G.-L. {Ingold}, Acta Phys. Pol. B {\bf 37}, 1537 (2006).

\bibitem{Ingold:2002pc}
G.-L. {Ingold}, {\em {Path Integrals and Their Application to
Dissipative Quantum Systems}\/}, Lect. Notes Phys.  {\bf 611}, 1
(2002).

\bibitem{Theo2002}
T.~M.~ Nieuwenhuizen and A.~E.~ Allahverdyan, Phys. Rev. E {\bf 66},
036102 (2002).


\bibitem{HaenggiQTransp98}
T.~Dittrich, P.~H{\"a}nggi, G.-L. Ingold, B.~Kramer, G.~Sch{\"o}n,
and W.~Zwerger, {\em Quantum Transport and Dissipation\/}
(Wiley-VCH, Weinheim, 1998).

\bibitem{Grabert:1988et}
H.~{Grabert}, P.~{Schramm}, and G.-L. {Ingold}, Phys. Rep. {\bf 168}, 115 (1988).


\bibitem{Grabert:1984ux}
H.~{Grabert}, U.~{Weiss}, and P.~{Talkner},  Z. Phys. B {\bf 55}, 87
(1984).


\bibitem{Caldeira:1983aj}
A.~O. {Caldeira} and A.~J. {Leggett}, Ann. Phys. (N.Y.) {\bf 149}, 374 (1983).


\bibitem{FeynStatMech}
R.~P. {Feynman}, {\em Statistical mechanics\/} (Addison Wesley, Redwood City,
  CA, 1972).

\bibitem{Feynman:1963qm}
R.~P. {Feynman} and F.~L. {Vernon}, Jr., Ann. Phys. (N.Y.)
  {\bf 24}, 118 (1963).




\end{thebibliography}
\end{document}